\documentclass[a4paper]{article}

\usepackage{INTERSPEECH2022}
\usepackage{amsmath,graphicx,underscore,booktabs,multirow,hyperref,cite,bm,enumitem}
\usepackage[safe]{tipa}
\hypersetup{hidelinks}
\title{Pronunciation Dictionary-Free Multilingual Speech Synthesis by Combining Unsupervised and Supervised Phonetic Representations}
\name{Chang Liu$^1$, Zhen-Hua Ling$^1$, Ling-Hui Chen$^2$ \thanks{This work was supported in part by the National Key R\&D Program of China under Grant 2019YFF0303001, and in part by the National Nature Science Foundation of China under Grant 61871358.}}
\address{
  $^1$NERC-SLIP, University of Science and Technology of China, Hefei, China \\
  $^2$iFLYTEK Research, Hefei, China}
\email{lcaily@mail.ustc.edu.cn, zhling@ustc.edu.cn, lhchen@iflytek.com}

\begin{document}

\maketitle
\begin{abstract}
This paper proposes a multilingual speech synthesis method which combines unsupervised phonetic representations (UPR) and supervised phonetic representations (SPR) to avoid reliance on the pronunciation dictionaries of target languages. 
In this method, a pretrained wav2vec 2.0 model is adopted to extract UPRs and a language-independent automatic speech recognition (LI-ASR) model is built with a connectionist temporal classification (CTC) loss to extract segment-level SPRs from the audio data of target languages. 
Then, an acoustic model is designed, which first predicts UPRs and SPRs from texts separately and then combines the predicted UPRs and SPRs to generate mel-spectrograms. 
The results of our experiments on six languages show that the proposed method outperformed the methods that directly predicted mel-spectrograms from character or phoneme sequences and the ablated models that utilized only UPRs or SPRs.
\end{abstract}

\section{Introduction}
\label{sec:intro}

Speech synthesis, also known as text-to-speech (TTS), aims to synthesize natural and intelligible speech from input text \cite{taylor2009text}. A TTS system is typically composed of a front-end and a back-end \cite{kumar2005human}. The front-end converts a text sequence into linguistic features, including several functionalities such as text normalization, word segmentation, part-of-speech (POS) tagging, prosody prediction and grapheme-to-phoneme (G2P) conversion \cite{tan2021survey}. The purpose of G2P conversion is to generate phoneme sequences from character sequences \cite{chae2018convolutional}. A pronunciation dictionary, which consists of word-pronunciation pairs of a language \cite{deri2016grapheme}, is essential for G2P conversion. 
However, pronunciation dictionaries are language-specific, and building a dictionary for a new language is labor-intensive, time-consuming and more difficult than acquiring speech recordings of the language \cite{deri2016grapheme}. Although there are some open source multilingual G2P tools such as Phonemizer \cite{Bernard2021}, the number of covered languages is still limited considering that there are approximately 7,000 languages around the world \cite{pereltsvaig2020languages}.

On the other hand, the back-end of a TTS system generally consists of an acoustic model, which converts linguistic features into acoustic features, and a vocoder to reconstruct speech from acoustic features. Currently, neural network-based sequence-to-sequence acoustic modeling \cite{Wang2017TacotronTE, shen2018natural, ren2019fastspeech, ren2020fastspeech, li2019neural} has become a mainstream approach. 
Some acoustic models can directly take character sequences as input \cite{Wang2017TacotronTE, shen2018natural}.
This benefits multilingual speech synthesis since pronunciation dictionaries and G2P conversion are not necessary anymore.
However, using character sequences as input usually degrades the naturalness and intelligibility of synthetic speech comparing with using phoneme sequences.

In addition to using phonemes to represent pronunciations, there are already some unsupervised models for learning pronunciation-related representations, such as vector-quantized variational autoencoder (VQ-VAE) \cite{van2017neural}, contrastive predictive coding (CPC) \cite{oord2018representation}, and wav2vec 2.0 \cite{baevski2020wav2vec}. These unsupervised phonetic representations have been applied to unsupervised speech recognition \cite{baevski2021unsupervised}, style control in speech synthesis \cite{chen2021fine}, speech resynthesis and voice conversion \cite{polyak2021speech}.
Although these unsupervised phonetic representations (UPR) contain pronunciation information, they differ significantly in granularity from phonemes and often include extra information such as prosody.

In this paper, we propose to extract supervised phonetic representations (SPR) and combine them with UPRs for multilingual speech synthesis method when pronunciation dictionaries of target languages are not available. 
Here,  UPRs are extracted using a pretrained wav2vec 2.0 model. 
The proposed SPRs are extracted using a language-independent automatic speech recognition (LI-ASR) model, which is supervisedly trained with a connectionist temporal classification (CTC) loss \cite{graves2006connectionist}.
Comparing with UPRs, one advantage of SPRs is that they are at segment-level and have similar granularity to phonemes. 
Besides, SPRs focus on describing pronunciation rather than prosody due to the supervised phoneme classification task of ASR. 
Comparing with using the discrete phonetic symbols recognized by LI-ASR, using SPRs can effectively mitigate the influence of recognition errors. 
By combining UPRs and SPRs, we expect to leverage the advantages of both representations. 
Given the speech corpus of a target language, we first extract UPRs and SPRs from speech waveforms, and then utilized them as intermediate representations to build the acoustic model.
Our proposed acoustic model includes two parts. First, two text-to-representation (TTR) models predict UPRs and SPRs from character sequences separately. Then, a representation-to-mel-spectrogram (RTM) model combines UPRs and SPRs to generate mel-spectrograms. 
Our experiments were conducted on six target languages. The results show that our proposed method significantly decreased the pronunciation errors and improved the naturalness of synthetic speech comparing with the method directly predicting mel-spectrograms from character or phoneme sequences. Besides, it also outperformed the ablated models that employed only URPs or SPRs.

\section{Proposed Method}
\label{sec:method}

In our proposed method, UPRs are extracted using a pretrained wav2vec 2.0 model following previous work \cite{baevski2021unsupervised}. In this section, we first introduce how to extract SPRs, and then describe the acoustic model combining UPRs and SPRs.

\begin{figure}[t]
	\centering
	\centerline{\includegraphics[width=0.9\linewidth]{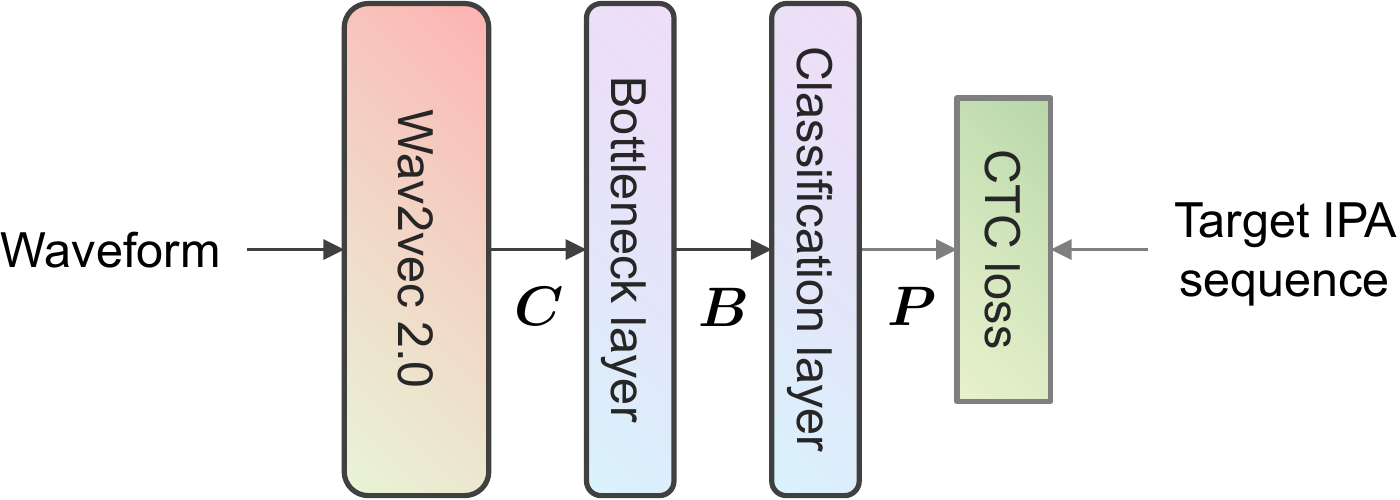}}
	\vspace{-0.2cm}
	\caption{The architecture of our LI-ASR model.}
	\label{fig:asr}
	\vspace{-0.5cm}
\end{figure}

\subsection{Extracting supervised phonetic representations (SPR) }
\label{ssec:asr}
SPRs are extracted using an LI-ASR model, 
which is trained to convert speech waveforms of different languages into the phoneme sequences with a shared
International Phonetic Alphabet (IPA) \cite{international1999handbook} phoneset. In this paper, wav2vec 2.0 \cite{baevski2020wav2vec} is adopted as the base architecture of the LI-ASR model. Wav2vec 2.0 is a self-supervised pretraining model that learns representations directly from speech waveforms by designing a contrastive task with unlabeled data. The pretrained wav2vec 2.0 model can be used for downstream speech recognition tasks to achieve better performance with less labeled data.
Fig.~\ref{fig:asr} illustrates the architecture of our LI-ASR model based on wav2vec 2.0. It adds two linear layers after the context representations and removes the quantization module in the original wav2vec 2.0 model. The first linear layer is called the bottleneck layer, which maps the 1024-dimensional context representations ($\bm{C}$) to 512-dimensional bottleneck representations ($\bm{B}$), from which SPRs are derived. The second layer is a classification layer, which predicts category probabilities ($\bm{P}$) from the bottleneck representations. The number of categories depends on the phoneset size. To learn shared phonetic representations across languages, the LI-ASR model is trained using a multilingual corpus with IPA phoneme transcriptions, and to minimize the CTC loss between the output category probabilities and the target IPA sequences. We also considered the Transformer-based decoder. However, its performance on unseen languages was not as good as the model structure introduced above in our preliminary experiments.

\begin{figure}[t]
  \centering
  \centerline{\includegraphics[width=0.9\linewidth]{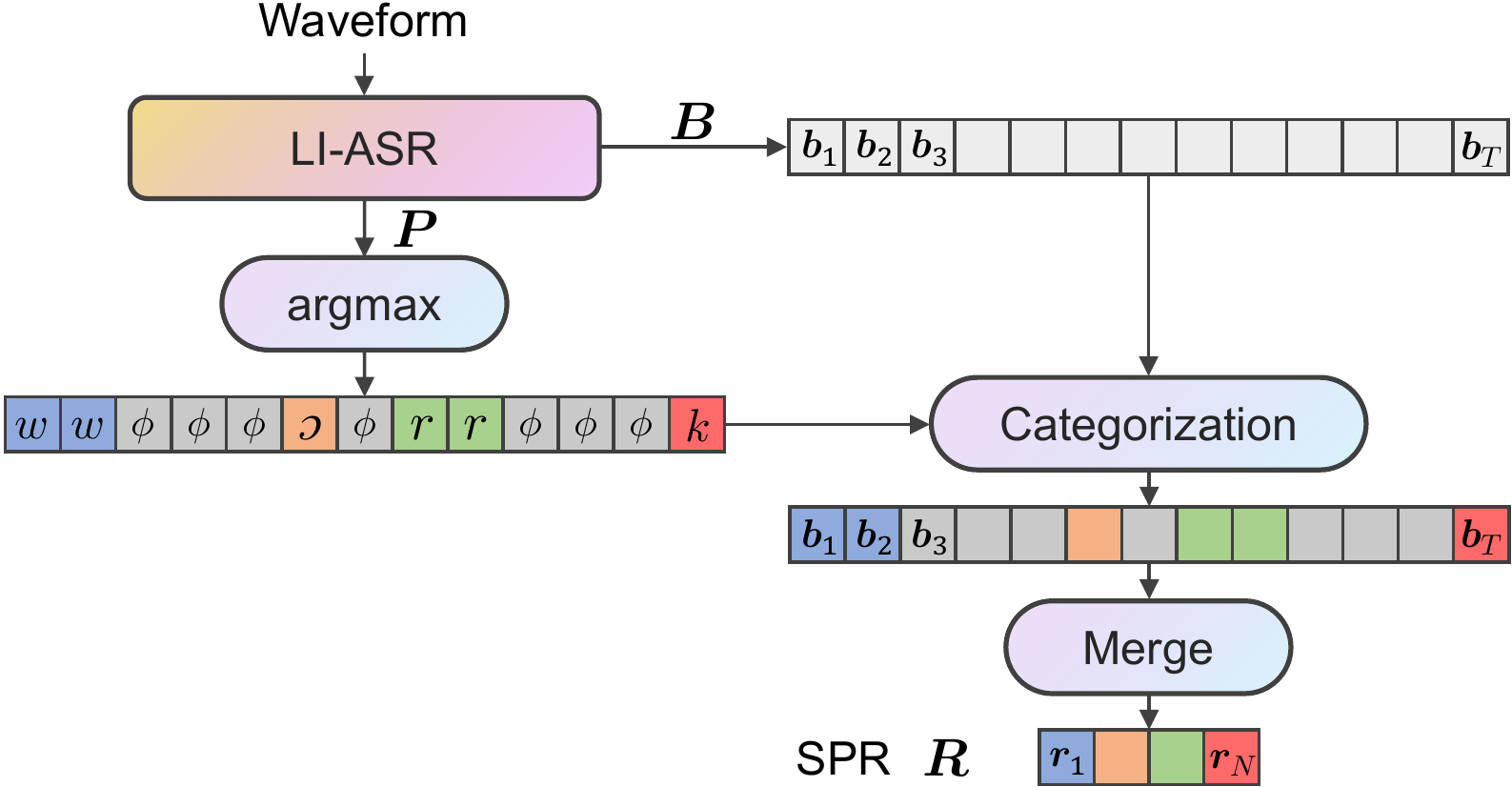}}
  \vspace{-0.2cm}
  \caption{The process of SPR extraction. Here, the audio clip of an English word \emph{``work"} is taken as an example. $\phi$ denotes the \emph{blank} symbol in CTC loss. $w$, \emph{\textopeno}, $r$ and $k$ are recognized phonetic symbols.}
  \label{fig:phore}
\end{figure}

Fig.~\ref{fig:phore} shows the process of extracting SPRs from the waveforms of a target language using the built LI-ASR model. Frame-level bottleneck representations $\bm{B}=\left[\bm{b}_{1}, \cdots, \bm{b}_{T}\right]$ are first calculated using the LI-ASR model, where $T$ is the frame number of the waveforms. Next, the $argmax$ operation is applied to category probabilities $\bm{P}$, yielding the corresponding category of each frame (a phonetic symbol or the blank symbol in CTC). Then, a \emph{Categorization} operation is conducted, which allocates the category of the $t$-th frame to the bottleneck representation $\bm{b}_t$. Finally, a \emph{Merge} operation is applied which removes the bottleneck representations with the \emph{blank} category and averages the adjacent bottleneck representations with same category into one vector, i.e., a SPR. In Fig.~\ref{fig:phore}, the extracted SPRs are written as $\bm{R}=\left[\bm{r}_{1}, \cdots, \bm{r}_{N}\right]$, where $N$ is the number of SPRs in the waveform. Obviously, $N$ is less than $T$.

\begin{figure}[t]
	\centering
	\centerline{\includegraphics[width=0.8\linewidth]{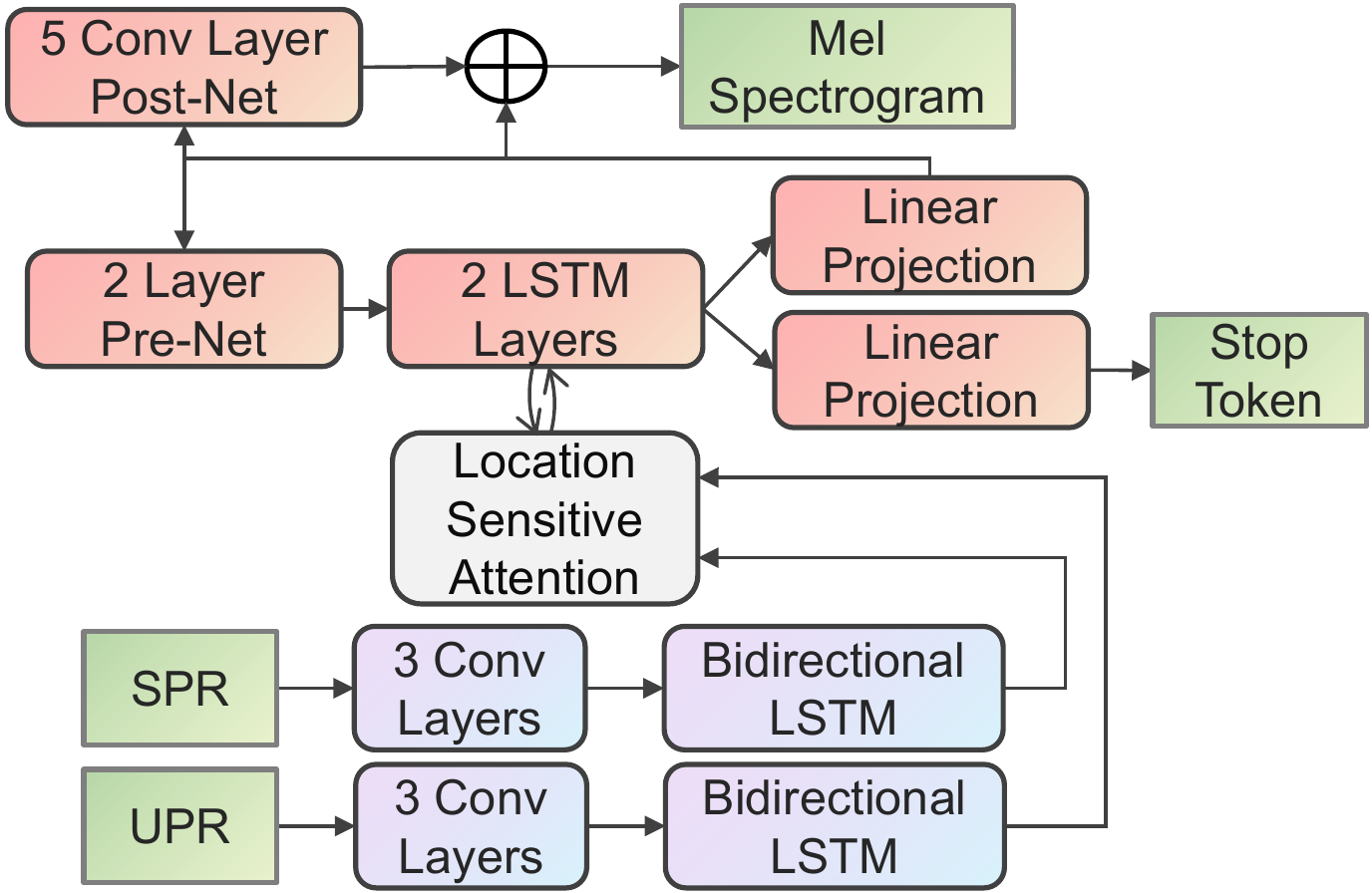}}
	\vspace{-0.2cm}
	\caption{The architecture of our RTM model.}
	\label{fig:mix}
	\vspace{-0.7cm}
\end{figure}

\subsection{Combining UPRs and SPRs for acoustic modeling}
\label{ssec:tts}

The acoustic model utilizes UPRs and SPRs as intermediate representations, and 
is composed of two text-to-representation (TTR) models together with a representation-to-mel-spectrogram (RTM) model. 
The two TTR models predict UPRs or SPRs from character sequences respectively.
Their architectures follow Tacotron2 \cite{shen2018natural} but with three differences.
First, the training target is changed from 80-dimensional mel-spectrograms to 512-dimensional UPRs or SPRs. Second, since the temporal continuity of UPRs and SPRs is not as obvious as acoustic features, the post-net is removed from the TTR models. Last, the dropout in pre-net is not applied during inference according to the results of our preliminary experiments. The loss function is composed of the mean squared error (MSE) loss and the L1 loss between the predicted UPRs or SPRs and the extracted references, together with the binary cross entropy (BCE) loss of stop tokens. 

As shown in Fig.~\ref{fig:mix}, the RTM model aims to generate mel-spectrograms by combining UPRs and SPRs. It has similar architecture to Tacotron2, except that it takes UPRs and SPRs as input rather than character or phoneme sequences. Thus, text embeddings are not required. First, two encoders with convolutional and bidirectional LSTM layers are used to process UPRs and SPRs simultaneously. 
Next, the location sensitve attention module computes context vectors using the outputs of the two encoders respectively and then averages them for decoding. At the top of Fig.~\ref{fig:mix}, it uses the same decoder architecture as Tacotron2. Its loss function consists of the MSE loss and L1 loss between the predicted mel-spectrograms and the real ones, and the BCE loss of stop tokens. 
At the synthesis stage, the predicted mel-spectrograms are sent into a HiFi-GAN vocoder to reconstruct speech waveforms.


\section{Experiments}
\label{sec:exp}

\subsection{Datasets}
\label{ssec:dataset}

We evaluated the performance of our proposed method on six target languages, i.e., \emph{English (en)}, \emph{Spanish (es)}, \emph{Kazakh (kk)}, \emph{Hindi (hi)}, \emph{Bulgarian (bg)}, and \emph{Malay (ms)}. The LJSpeech dataset \cite{ljspeech17} was used for English and proprietary single speaker speech synthesis datasets were used for all other languages. The speakers of all languages were female. 
We set the training set size for each language to the same 5 hours, corresponding to 2712, 4356, 3013, 3031, 3006 and 2588 utterances respectively. 
Both the validation set and the test set contained 300, 400, 200, 250, 150 and 200 utterances respectively. 
All utterances in the training, validation and test sets had recordings with corresponding text transcriptions, and  there were no overlapped utterances between these three sets. 
Besides, we used another test set with only text for ASR-based intelligibility evaluation, which contained 1000 sentences for each language. 

In order to build the LI-ASR model, the multilingual wav2vec 2.0 model provided  by Facebook was adopted, which was pretrained using 53 languages\footnote{\url{https://github.com/pytorch/fairseq/tree/main/examples/wav2vec}}. To fine-tune the LI-ASR model, we used another proprietary corpus of 19 languages, including \emph{Bengali (bn)},  \emph{Catalan (ca)},  \emph{Czech (cs)},  \emph{German (de)},  \emph{Greek (el)},  \emph{French (fr)},  \emph{Hungarian (hu)},  \emph{Indonesian (id)},  \emph{Italian (it)},  \emph{Norwegian (nb)},  \emph{Dutch (nl)},  \emph{Polish (pl)},  \emph{Portuguese (pt)},  \emph{Russian (ru)},  \emph{Slovenian (sl)},  \emph{Serbian (sr)},  \emph{Swedish (sv)},  \emph{Turkish (tr)},  \emph{Urdu (ur)}. 
The duration of each language was 100 to 150 hours, contributing to total 2771 hours. 
Each utterance had a corresponding text transcription. Note that there were no overlapped languages between the fine-tuning corpus and the synthesis corpus. Although English, Spanish and Kazakh were included in the pretraining corpus of wav2vec 2.0, no text or phonetic labels of these languages were utilized.

All speech recordings used in our experiments were sampled at 16 kHz, otherwise downsampled to 16 kHz.

\subsection{Experimental setup}
\label{ssec:setup}

For fine-tuning the LI-ASR model, we employed the open source tool Phonemizer \cite{Bernard2021} to convert the transcriptions in the corpus of 19 languages into IPA phoneme sequences without punctuations. The size of the obtained IPA phoneset was 203, thus the output dimension of the classification layer in the LI-ASR model was 204 with an extra CTC \emph{blank} symbol. 
The other configurations of the LI-ASR model were the same as the \textsc{Large} model in previous work \cite{baevski2020wav2vec}. 
The LI-ASR model was implemented based on the fairseq framework \cite{ott2019fairseq}. Finally, we chose the model checkpoint with the lowest phoneme error rate (PER) of 5.87\% on the validation sets of 19 languages.

We compared our proposed method with five methods.\footnote{Audio samples are available at \url{https://ryuclc.github.io/SPR}.} 
\begin{itemize}[itemsep=1pt,topsep=1pt,parsep=0pt]
    \item \textbf{Taco-Char} A Tacotron2 model using character sequences as input. This was the baseline model.
    \item \textbf{Taco-Phone} A Tacotron2 model using phoneme sequences as input. We used Phonemizer to convert the transcriptions of target languages into phoneme sequences except using Festival\footnote{\url{https://www.cstr.ed.ac.uk/projects/festival}} for English.
    \item \textbf{DPS} It adopted the discrete phonetic symbols (DPS) recognized by the LI-ASR model as intermediate representations instead of UPRs and SPRs. Here, the TTR model was a G2P model based on long short-term memory (LSTM) with CTC, and the RTM model took the predicted DPSs, i.e., phoneme sequences, as input.
    \item \textbf{SPR/UPR} Two ablated methods with only SPRs and UPRs. Only one TTR model was built to predict either SPRs or UPRs, and the RTM model accepted either SPRs or UPRs as inputs.
\end{itemize}

In our implementation, 80-dimensional mel-spectrograms were used as acoustic features. The configurations of Tacotron2 followed its original paper \cite{shen2018natural}. We chose the model checkpoint with the lowest loss on the validation set. HiFi-GAN vocoder \cite{kong2020hifi} was employed to generate speech waveforms from mel-spectrograms. Note that we trained separate speech synthesis models and vocoders for different languages.

The character error rate (CER) given by ASR was adopted to evaluate the intelligibility of synthetic speech. We used Google speech API to transcribe the synthetic speech for CER calculation. As mentioned above, a test set of 1000 sentences was adopted in the CER evaluation for each language. 
Our subjective evaluation included mean opinion score (MOS) tests and preference listening tests to evaluate the naturalness of synthetic speech. For each language, 30 sentences were selected from the test set and were synthesized using different methods. In MOS tests, the naturalness of each synthetic utterance was scored in a scale from 1 to 5 with an interval of 0.5 by native speakers for each target language. In preference listening tests, native speakers were asked to listen to paired synthetic utterances and then give their preference on naturalness for each pair.

\subsection{Experimental results}
\label{ssec:result}

\begin{table}[t]
	\centering
	\caption{The CERs (\%) of different methods on six languages.}
	\vspace{-0.2cm}
	\label{tab:CER5h}
	\begin{tabular}{ccccccc}
		\toprule
		\specialrule{0em}{0pt}{0pt}
		& en & es & kk & hi & bg & ms \\ [-3pt]
		\midrule
		\specialrule{0em}{0pt}{0pt}
		Taco-Char        & 15.61 & 1.64 & 7.19 & 3.25 & 3.45 & 2.21 \\
		Taco-Phone      & 6.64 & 1.60 & 6.85 & 1.74 & 3.72 & 2.32 \\
		DPS                 & 29.13 & 4.11 & 19.87 & 12.45 & 8.37 & 3.56 \\
		SPR           & 6.31 & 1.60 & 5.78 & 1.89 & 3.33 & 2.01 \\
		UPR           & 6.36 & 1.56 & 5.63 & 1.90 & 3.94 & 1.96 \\
		Proposed   & \textbf{5.97} & \textbf{1.33} & \textbf{5.47} & \textbf{1.58} & \textbf{3.07} & \textbf{1.81} \\ [-3pt]
		\bottomrule
	\end{tabular}
	\vspace{-0.5cm}
\end{table}

\begin{table*}[t]
	\centering
	\caption{The naturalness MOS of different methods on six target languages with 95\% confidence intervals.}
	\vspace{-0.2cm}
	\label{tab:MOS}
	\begin{tabular}{ccccccc}
		\toprule
		\specialrule{0em}{0pt}{0pt}
		& en & es & kk & hi & bg & ms \\ [-3pt]
		\midrule
		\specialrule{0em}{0pt}{0pt}
		Ground truth & 4.26 $\pm$ 0.10 & 4.64 $\pm$ 0.07 & 4.87 $\pm$ 0.04 & 4.51 $\pm$ 0.08 & 4.88 $\pm$ 0.05 & 4.29 $\pm$ 0.09 \\
		Taco-Char   & 2.89 $\pm$ 0.16 & 4.42 $\pm$ 0.07 & 4.32 $\pm$ 0.07 & 4.31 $\pm$ 0.08 & 4.15 $\pm$ 0.13 & 3.90 $\pm$ 0.08 \\
		Taco-Phone  & 3.19 $\pm$ 0.15 & 4.33 $\pm$ 0.07 & 4.40 $\pm$ 0.07 & 4.41 $\pm$ 0.07 & 4.13 $\pm$ 0.12 & 3.91 $\pm$ 0.09 \\
		Proposed   & \textbf{3.89 $\pm$ 0.13} & \textbf{4.52 $\pm$ 0.07} & \textbf{4.67 $\pm$ 0.07} & 4.43 $\pm$ 0.07 & \textbf{4.39 $\pm$ 0.10} & \textbf{4.16 $\pm$ 0.08} \\ [-3pt]
		\bottomrule
	\end{tabular}
	\vspace{-0.6cm}
\end{table*}

Table \ref{tab:CER5h} shows the CER results. 
It is obvious that the DPS method performed worst due to the errors caused by the hard decisions when recognizing discrete phonemes. 
Taco-Phone performed better than Taco-Char on four languages, but worse on \emph{bg} and \emph{ms}. One possible reason is that the Phonemizer tool may not provide accurate enough phoneme transcriptions for  \emph{bg} and \emph{ms}. 
SPR and UPR achieved similar CERs, and they both outperformed Taco-Char except that UPR performed worse than Taco-Char on \emph{bg}. 
Comparing with Taco-Phone, SPR achieved lower CERs on 5 languages and UPR achieved lower CERs on 4 languages. 
These results demonstrate the effectiveness of using UPRs or SPRs on improving the intelligibility of synthetic speech when the phoneme transcriptions of target languages are not available.
Furthermore, our proposed method achieved the lowest CERs among all compared methods for all six language, which shows the advantage of combining SPRs and UPRs for acoustic modeling.

\begin{table}[t]
	\centering
	\caption{Preference scores (\%) on naturalness between the proposed method and two ablated methods on six target languages, where N/P denotes ``No preference" and $p$ means the $p$-value of paired $t$-test between two methods.}
	\vspace{-0.2cm}
	\label{tab:PRE}
	\begin{tabular}{c|ccccc}
		\toprule
		\specialrule{0em}{0pt}{0pt}
		& Proposed &       SPR       &  UPR    &     N/P    &    $p$    \\ [-3pt]
		\midrule
		\specialrule{0em}{0pt}{0pt}
		\multirow{2}{*}{en}      &\textbf{51.11} &   21.11     &         -      &      27.78     & $<$0.001  \\
											 &23.33 &         -       &      19.44    &      57.22     &  0.43   \\ [-3pt]
		\midrule
		\specialrule{0em}{0pt}{0pt}       
		\multirow{2}{*}{es}     &35.83 &      23.33     &         -         &      40.83     & 0.07   \\
										    &3.33     &             -         &      3.33    &      93.33     &  1.00   \\ [-3pt]
		\midrule
		\specialrule{0em}{0pt}{0pt}                              
		\multirow{2}{*}{kk}     &\textbf{43.33} &      20.56     &         -         &      36.11     & $<$0.001   \\
										   &12.67     &             -         &      12.00    &      75.33     &  0.87   \\ [-3pt]
		\midrule
		\specialrule{0em}{0pt}{0pt} 
		\multirow{2}{*}{hi}      &35.33 &      35.33     &         -         &      29.33     & 1.00   \\
										   & 5.00     &             -         &      5.83    &      89.17     &  0.78   \\ [-3pt]
		\midrule
		\specialrule{0em}{0pt}{0pt} 
		\multirow{2}{*}{bg}      &28.67 &      18.67     &         -         &       52.67     & 0.07   \\
										    & \textbf{27.33}     &             -         &      16.67    &     56.00     &  0.04   \\ [-3pt]
		\midrule
		\specialrule{0em}{0pt}{0pt} 
		\multirow{2}{*}{ms}      & \textbf{62.22} &      18.33     &         -         &    19.44     & $<$0.001   \\
											 & 15.33     &             -         &   19.33    &     65.33     &  0.41   \\ [-3pt]						
		\bottomrule
	\end{tabular}
	\vspace{-0.3cm}
\end{table}

\begin{table}[t]
	\centering
	\caption{The LD and LMR (\%) of extracted UPRs and SPRs. More details on the metrics can be found in Section \ref{ssec:analysis}. }
	\vspace{-0.2cm}
	\label{tab:LD}
	\renewcommand\tabcolsep{4pt} 
	\begin{tabular}{c|ccccccc}
		\toprule
		\specialrule{0em}{0pt}{0pt}
		& & en & es & kk & hi & bg & ms \\ [-3pt]
		\midrule
		\specialrule{0em}{0pt}{0pt}
		\multirow{2}{*}{LD}  &  SPR       & \textbf{2.29} & \textbf{1.19} & \textbf{6.06} & \textbf{1.54} & \textbf{1.81} & \textbf{2.01} \\
		&  UPR      & 33.92 & 31.97 & 31.37 & 38.75 & 29.50 & 35.91 \\
		\hline
		\specialrule{0em}{0pt}{0pt} 
		\multirow{2}{*}{LMR}  &  SPR       & \textbf{3.63} & \textbf{1.94} & \textbf{6.71} & \textbf{2.73} & \textbf{2.45} & \textbf{2.60} \\
		&  UPR      & 51.99 & 55.05 & 39.42 & 63.03 & 39.82 & 47.38 \\ [-3pt]
		\bottomrule
	\end{tabular}
	\vspace{-0.4cm}
\end{table}

Due to the poor CER performance of the DPS model, it was not included in our subjective evaluations. 
The MOS results of comparing our proposed method with Taco-Char and Taco-Phone are shown in Table \ref{tab:MOS}. 
Here, the utterances reconstructed from ground truth mel-spectrograms by the HiFi-GAN vocoder were used for reference.
The $p$-value of paired $t$-test was used to measure the significance of the difference between two methods. 
Similar to the results in Table \ref{tab:CER5h}, Taco-Phone achieved better naturalness than Taco-Char on \emph{en, kk, hi} ($p=2\times10^{-3}, 0.03, 6\times10^{-3}$). However, it was comparable with Taco-Char on \emph{bg} ($p=0.66$) and \emph{ms} ($p=0.77$), and even worse than Taco-Char on \emph{es} ($p=0.01$). One possible reason is that the Phonemizer tool may not provide accurate enough phoneme transcriptions for the last three languages. 
Our proposed method performed significantly better than Taco-Char on \emph{en}, \emph{es}, \emph{kk}, \emph{hi}, \emph{bg} and \emph{ms} ($p=1\times10^{-17}, 0.01, 3\times10^{-11}, 5\times10^{-4}, 2\times10^{-5}, 1\times10^{-9}$). Besides, compared with Taco-Phone, our proposed model achieved significantly better naturalness on \emph{en}, \emph{es}, \emph{kk}, \emph{bg} and \emph{ms} ($p=1\times10^{-10}, 3\times10^{-5}, 2\times10^{-7}, 4\times10^{-8}, 2\times10^{-8}$) and they were comparable on \emph{hi} ($p=0.67$). These results also demonstrate the superiority of our proposed method.

Preference listening tests were conducted to compare the naturalness of our proposed method with the two ablated methods.
The results are listed in Table~\ref{tab:PRE}. 
We can see that our proposed method performed better than SPR on three languages and better than UPR on one language. On the other languages, the preference differences were insignificant. This shows the benefit of combining UPRs and SPRs on synthesizing more natural speech, in addition to improving intelligibility.

\subsection{Analysis on phonetic representations}
\label{ssec:analysis}
The proposed SPRs are expected to have two characteristics. First, they are segment-level ones with similar granularity as phonemes. Second, they can carry rich pronunciation information of phonemes. 
Therefore, two additional experiments were conducted to examine whether the extracted SPRs were in line with our expectations.

We first investigated the granularity of extracted SPRs. 
Two metrics were designed for evaluation, length difference (LD) and length mismatch rate (LMR). 
Assume that there are $M$ test utterances, and $L_{SPR}^{(m)}$ and $L_{IPA}^{(m)}$ denote the number of extracted SPRs and the number of IPA phonemes generated by Phonemizer for the $m$-th utterance. 
Then, LD is defined as
\begin{equation}
\setlength{\abovedisplayskip}{3pt}
\setlength{\belowdisplayskip}{3pt}
	\label{equ:LD}
	LD=\sum_{m}{\left|L_{SPR}^{(m)}-L_{IPA}^{(m)}\right|}/M\text{,}
\end{equation}
and LMR is defined as
\begin{equation}
\setlength{\abovedisplayskip}{3pt}
\setlength{\belowdisplayskip}{3pt}
  \label{equ:LMR}
  LMR=\sum_{m}\frac{\left|L_{SPR}^{(m)}-L_{IPA}^{(m)}\right|}{L_{IPA}^{(m)}}/M\text{.}
\end{equation}
Similarly, we also computed the two metrics of UPRs.
Table \ref{tab:LD} shows the results on the six target languages. We can see that the extracted SPRs had similar granularity to phonemes on all six target languages. But UPRs differed significantly from phonemes in terms of granularity. 

%
\begin{figure}[t]
	\centering
	\centerline{\includegraphics[width=0.97\linewidth]{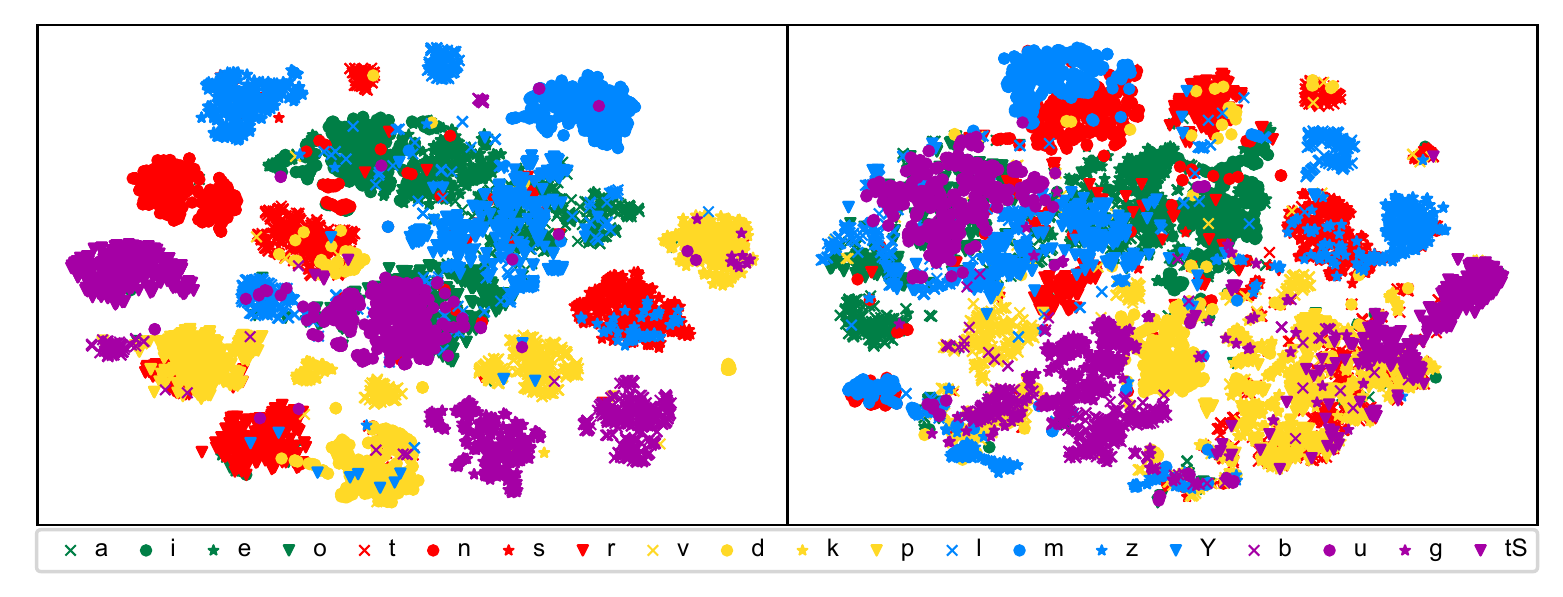}}
	\vspace{-0.1cm}
	\caption{t-SNE visualization of SPRs (left) and UPRs (right) from the Bulgarian speech synthesis dataset.}
	\label{fig:bg}
\vspace{-0.4cm}
\end{figure}

We further investigated the mapping relationship between the extracted SPRs or UPRs and phonemes.
To obtain the phoneme categories of both representations, Montreal Forced Aligner (MFA) \cite{mcauliffe2017montreal} was used to align phoneme sequences to audio. A phoneme category was assigned to a representation if the duration of the phoneme covered the entire receptive field of the representation. Visualization analysis was conducted on the speech synthesis datasets of English and Bulgarian by t-Distributed Stochastic Neighbor Embedding (t-SNE)\footnote{\url{https://lvdmaaten.github.io/tsne}}. We randomly selected 500 representations for each of the 20 most frequent phonemes in each language. The phonesets of MFA for English and Bulgarian are ARPAbet and GlobalPhone, respectively.
The  results of Bulgarian are shown in Fig.~\ref{fig:bg} and the results of English are available at our demo webpage due to limited space.
In both figures, we can see that SPRs showed better phoneme-dependent clustering property than UPRs. 
This indicates that the supervised training of the LI-ASR model helps to enhance the pronunciation-relatedness of extracted SPRs.

\section{Conclusion}
\label{sec:conclusion}
This paper has proposed a multilingual speech synthesis method without reliance on pronunciation dictionaries.
In addition to the unsupervised phonetic representations (UPR) extracted by a pretrained wav2vec 2.0 model, we propose to extract supervised phonetic representations (SPR) using an LI-ASR model. SPRs are more similar to phonemes in granularity and are more pronunciation-related than UPRs, while UPRs contain richer information besides pronunciation. Our proposed method leverages the advantages of both UPRs and SPRs by combining them in acoustic modeling. Experimental results have demonstrated the effectiveness of our proposed method on improving the intelligibility and naturalness of synthetic speech.
Future work includes finding a better architecture of LI-ASR and developing an acoustic model shared across languages.

\bibliographystyle{IEEEtran}

\bibliography{mybib}

\begin{thebibliography}{10}
\providecommand{\url}[1]{#1}
\csname url@samestyle\endcsname
\providecommand{\newblock}{\relax}
\providecommand{\bibinfo}[2]{#2}
\providecommand{\BIBentrySTDinterwordspacing}{\spaceskip=0pt\relax}
\providecommand{\BIBentryALTinterwordstretchfactor}{4}
\providecommand{\BIBentryALTinterwordspacing}{\spaceskip=\fontdimen2\font plus
\BIBentryALTinterwordstretchfactor\fontdimen3\font minus
  \fontdimen4\font\relax}
\providecommand{\BIBforeignlanguage}[2]{{%
\expandafter\ifx\csname l@#1\endcsname\relax
\typeout{** WARNING: IEEEtran.bst: No hyphenation pattern has been}%
\typeout{** loaded for the language `#1'. Using the pattern for}%
\typeout{** the default language instead.}%
\else
\language=\csname l@#1\endcsname
\fi
#2}}
\providecommand{\BIBdecl}{\relax}
\BIBdecl

\bibitem{taylor2009text}
P.~Taylor, \emph{Text-to-speech synthesis}.\hskip 1em plus 0.5em minus
  0.4em\relax Cambridge university press, 2009.

\bibitem{kumar2005human}
R.~Kumar, \emph{Human computer interaction}.\hskip 1em plus 0.5em minus
  0.4em\relax Firewall Media, 2005.

\bibitem{tan2021survey}
X.~Tan, T.~Qin, F.~Soong, and T.-Y. Liu, ``A survey on neural speech
  synthesis,'' \emph{arXiv preprint arXiv:2106.15561}, 2021.

\bibitem{chae2018convolutional}
M.-J. Chae, K.~Park, J.~Bang, S.~Suh, J.~Park, N.~Kim, and L.~Park,
  ``Convolutional sequence to sequence model with non-sequential greedy
  decoding for grapheme to phoneme conversion,'' in \emph{2018 IEEE
  International Conference on Acoustics, Speech and Signal Processing
  (ICASSP)}.\hskip 1em plus 0.5em minus 0.4em\relax IEEE, 2018, pp. 2486--2490.

\bibitem{deri2016grapheme}
A.~Deri and K.~Knight, ``Grapheme-to-phoneme models for (almost) any
  language,'' in \emph{Proceedings of the 54th Annual Meeting of the
  Association for Computational Linguistics (Volume 1: Long Papers)}, 2016, pp.
  399--408.

\bibitem{Bernard2021}
\BIBentryALTinterwordspacing
M.~Bernard and H.~Titeux, ``Phonemizer: Text to phones transcription for
  multiple languages in python,'' \emph{Journal of Open Source Software},
  vol.~6, no.~68, p. 3958, 2021. [Online]. Available:
  \url{https://doi.org/10.21105/joss.03958}
\BIBentrySTDinterwordspacing

\bibitem{pereltsvaig2020languages}
A.~Pereltsvaig, \emph{Languages of the World}.\hskip 1em plus 0.5em minus
  0.4em\relax Cambridge University Press, 2020.

\bibitem{Wang2017TacotronTE}
Y.~Wang, R.~Skerry-Ryan, D.~Stanton, Y.~Wu, R.~J. Weiss, N.~Jaitly, Z.~Yang,
  Y.~Xiao, Z.~Chen, S.~Bengio, Q.~V. Le, Y.~Agiomyrgiannakis, R.~Clark, and
  R.~Saurous, ``Tacotron: Towards end-to-end speech synthesis,'' in
  \emph{INTERSPEECH}, 2017.

\bibitem{shen2018natural}
J.~Shen, R.~Pang, R.~J. Weiss, M.~Schuster, N.~Jaitly, Z.~Yang, Z.~Chen,
  Y.~Zhang, Y.~Wang, R.~Skerrv-Ryan \emph{et~al.}, ``Natural {TTS} synthesis by
  conditioning wavenet on mel spectrogram predictions,'' in \emph{2018 IEEE
  International Conference on Acoustics, Speech and Signal Processing
  (ICASSP)}.\hskip 1em plus 0.5em minus 0.4em\relax IEEE, 2018, pp. 4779--4783.

\bibitem{ren2019fastspeech}
Y.~Ren, Y.~Ruan, X.~Tan, T.~Qin, S.~Zhao, Z.~Zhao, and T.-Y. Liu, ``Fastspeech:
  fast, robust and controllable text to speech,'' in \emph{Proceedings of the
  33rd International Conference on Neural Information Processing Systems},
  2019, pp. 3171--3180.

\bibitem{ren2020fastspeech}
Y.~Ren, C.~Hu, X.~Tan, T.~Qin, S.~Zhao, Z.~Zhao, and T.-Y. Liu, ``Fastspeech 2:
  Fast and high-quality end-to-end text to speech,'' in \emph{International
  Conference on Learning Representations}, 2020.

\bibitem{li2019neural}
N.~Li, S.~Liu, Y.~Liu, S.~Zhao, and M.~Liu, ``Neural speech synthesis with
  transformer network,'' in \emph{Proceedings of the AAAI Conference on
  Artificial Intelligence}, vol.~33, no.~01, 2019, pp. 6706--6713.

\bibitem{van2017neural}
A.~van~den Oord, O.~Vinyals, and K.~Kavukcuoglu, ``Neural discrete
  representation learning,'' in \emph{Proceedings of the 31st International
  Conference on Neural Information Processing Systems}, 2017, pp. 6309--6318.

\bibitem{oord2018representation}
A.~v.~d. Oord, Y.~Li, and O.~Vinyals, ``Representation learning with
  contrastive predictive coding,'' \emph{arXiv preprint arXiv:1807.03748},
  2018.

\bibitem{baevski2020wav2vec}
A.~Baevski, Y.~Zhou, A.~Mohamed, and M.~Auli, ``wav2vec 2.0: A framework for
  self-supervised learning of speech representations,'' \emph{Advances in
  Neural Information Processing Systems}, vol.~33, 2020.

\bibitem{baevski2021unsupervised}
A.~Baevski, W.-N. Hsu, A.~Conneau, and M.~Auli, ``Unsupervised speech
  recognition,'' \emph{Advances in Neural Information Processing Systems},
  vol.~34, 2021.

\bibitem{chen2021fine}
L.-W. Chen and A.~Rudnicky, ``Fine-grained style control in transformer-based
  text-to-speech synthesis,'' \emph{arXiv preprint arXiv:2110.06306}, 2021.

\bibitem{polyak2021speech}
A.~Polyak, Y.~Adi, J.~Copet, E.~Kharitonov, K.~Lakhotia, W.-N. Hsu, A.~Mohamed,
  and E.~Dupoux, ``Speech resynthesis from discrete disentangled
  self-supervised representations,'' in \emph{INTERSPEECH 2021-Annual
  Conference of the International Speech Communication Association}, 2021.

\bibitem{graves2006connectionist}
A.~Graves, S.~Fern{\'a}ndez, F.~Gomez, and J.~Schmidhuber, ``Connectionist
  temporal classification: labelling unsegmented sequence data with recurrent
  neural networks,'' in \emph{Proceedings of the 23rd international conference
  on Machine learning}, 2006, pp. 369--376.

\bibitem{international1999handbook}
I.~P. Association, I.~P.~A. Staff \emph{et~al.}, \emph{Handbook of the
  International Phonetic Association: A guide to the use of the International
  Phonetic Alphabet}.\hskip 1em plus 0.5em minus 0.4em\relax Cambridge
  University Press, 1999.

\bibitem{ljspeech17}
K.~Ito and L.~Johnson, ``The lj speech dataset,''
  \url{https://keithito.com/LJ-Speech-Dataset/}, 2017.

\bibitem{ott2019fairseq}
M.~Ott, S.~Edunov, A.~Baevski, A.~Fan, S.~Gross, N.~Ng, D.~Grangier, and
  M.~Auli, ``fairseq: A fast, extensible toolkit for sequence modeling,'' in
  \emph{Proceedings of NAACL-HLT 2019: Demonstrations}, 2019.

\bibitem{kong2020hifi}
J.~Kong, J.~Kim, and J.~Bae, ``Hifi-gan: Generative adversarial networks for
  efficient and high fidelity speech synthesis,'' \emph{Advances in Neural
  Information Processing Systems}, vol.~33, 2020.

\bibitem{mcauliffe2017montreal}
M.~McAuliffe, M.~Socolof, S.~Mihuc, M.~Wagner, and M.~Sonderegger, ``Montreal
  forced aligner: Trainable text-speech alignment using kaldi.'' in
  \emph{Interspeech}, vol. 2017, 2017, pp. 498--502.

\end{thebibliography}


\end{document}